# High efficiency laser-assisted H⁻ charge exchange for microsecond duration beams

Sarah Cousineau,[1,*] Abdurahim Rakhman,[1] Martin Kay,[1] Alexander Aleksandrov,[2]
Viatcheslav Danilov,[2,†] Timofey Gorlov,[2] Yun Liu,[2] Cary Long,[2] Alexander Menshov,[2]
Michael Plum,[2] Andrei Shishlo,[2] Andrew Webster,[2] and David Johnson[3]

[1]*Department of Physics and Astronomy, University of Tennessee, Knoxville, Tennessee 37966, USA*
[2]*Oak Ridge National Laboratory, Oak Ridge, Tennessee 37831, USA*
[3]*Fermilab National Laboratory, Batavia, Illinois 60510, USA*

(Received 8 August 2017; published 26 December 2017)

Laser-assisted stripping is a novel approach to H⁻ charge exchange that overcomes long-standing limitations associated with the traditional, foil-based method of producing high-intensity, time-structured beams of protons. This paper reports on the first successful demonstration of the laser stripping technique for microsecond duration beams. The experiment represents a factor of 1000 increase in the stripped pulse duration compared with the previous proof-of-principle demonstration. The central theme of the experiment is the implementation of methods to reduce the required average laser power such that high efficiency stripping can be accomplished for microsecond duration beams using conventional laser technology. The experiment was performed on the Spallation Neutron Source 1 GeV H⁻ beam using a 1 MW peak power UV laser and resulted in ~95% stripping efficiency.

## I. INTRODUCTION

Many accelerator applications require intense, time-structured beams of protons. The defacto method for producing these beams is through an H⁻ charge exchange injection from a linac into a synchrotron or accumulator ring. In this method, a magnet is used to merge an incoming H⁻ beam with a circulating proton beam in a ring. The merged beam is then stripped of its electrons to produce a single-species proton beam. This method was invented in 1951 [1] and has several advantages over direct proton injection, including high capture efficiency and minimum phase space growth [2].

While this non-Liousvillian approach can yield extremely high proton beam densities, in practice the achievable densities are limited by the conventional electron stripping mechanism. The nominal charge exchange configuration relies on a thin ($\mu g/cm^2$) foil of low-$Z$ material (such as carbon) to strip the two electrons from the H⁻ beam after a merger with the circulating proton beam. Unfortunately, the presence of the foil in the beam line introduces significant performance limitations. First, particle scattering in the foil leads to large levels of beam loss and activation, thus complicating the routine, hands-on maintenance of injection system components. Second, foils suffer vulnerabilities in their structural integrity when thermal effects are significant, reducing mean lifetimes. The primary failure mechanism for carbon foils is sublimation above a threshold temperature [3], which translates into constraints on the allowable beam power density. Though there is some uncertainty in the calculations, the practical upper limit on the beam power density is estimated to be within an order of magnitude of today's values, with some machines already operating on the lower end of the sublimation temperature range. Schemes such as injection painting can be used to minimize beam foil traversals and their subsequent effects but are not infinitely scalable, because they result in progressively larger beam emittances and machine apertures [4]. Currently, there is no viable alternative for replacing the foil-based system once this threshold is reached, thereby limiting the achievable beam parameters in future synchrotron-based proton accelerators.

The H⁻ laser stripping concept, originated over three decades ago [5], offers an attractive alternative that replaces the conventional foil-based configuration with a laser and dipole ensemble. In this material-free version of the charge exchange method, the first, loosely bound outer electron is stripped by a high field dipole magnet, converting H⁻ to H⁰. While in theory it is possible to remove the second electron using direct laser photodetachment, it would require excessively high peak laser powers. Instead, a laser is used to produce resonant excitation of the remaining electron to a higher quantum state (H⁰*) with a lower binding energy [6]. The excited electron is then Lorentz stripped by a second dipole magnet to produce a proton ($H^{0*}$ to $p$). The process is shown schematically in Fig. 1. This method is scalable to arbitrarily high beam power densities and completely relieves the issue of beam loss and radiation from particle scattering.

---
[*]Also at Oak Ridge National Laboratory, Oak Ridge, Tennessee 37831, USA.
mcousin2@utk.edu
[†]Deceased.

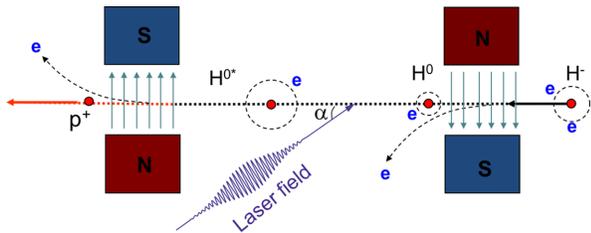

FIG. 1. Schematic of the laser stripping concept in this experiment, showing Lorentz stripping of the first electron by a dipole magnet in the first step (far right), resonant excitation of the second electron by the laser in the second step (middle); and finally stripping of the excited electron by the second dipole magnet (left). This figure is reproduced from Ref. [7].

For decades, the concept of laser stripping saw no experimental realization. This was due to a fundamental complication arising from the inherent energy spread in the ion beam, translating into a spread in the resonant excitation frequency of the ion beam particles, beyond the obtainable laser bandwidths. In 2006, this problem was overcome in a proof-of-principle (POP) experiment that utilized a diverging laser to induce a frequency sweep in the rest frame of the ion beam [8]. The POP experiment was conducted at the Spallation Neutron Source accelerator, a 1.4 MW, 1 GeV superconducting linear accelerator. The frequency sweep was achieved using a $Q$-switched UV laser with a 0.4° divergence angle. This configuration required 10 MW of peak laser power to provide the requisite laser power density for the high efficiency excitation of the intermediate $H^0$ beam to the $n = 3$ quantum state. The laser duty factor was 7 ns at 30 Hz, for an average laser power of about 2 W. The experiment successfully demonstrated 90% stripping of the 7 ns, 1 GeV $H^-$ beam and marked the start of the experimental evolution of laser stripping.

While the POP demonstration was a landmark accomplishment that validated the concept, the stripped beam pulse was still orders of magnitude shorter than a typical ion beam macropulse. A direct scaling of the POP experiment to the full Spallation Neutron Source ion beam pulse duty factor of 1 ms and 60 Hz yields a required average UV laser power of approximately 600 kW, whereas the current state-of-the-art laser technology is on the order of tens of watts. Thus, to extend the laser stripping method into the realm of practical ion pulse lengths, it is necessary to reduce the required average laser power to feasible levels. This can be accomplished by the clever manipulation of both the $H^-$ beam and laser beam parameters, as described by Ref. [8]. The power-saving methods cumulatively result in a 3 orders of magnitude reduction in the required average laser power, making it possible to strip microsecond duration beam pulses with conventional laser technology.

The goal of the present work was to implement the laser power-saving techniques to demonstrate high efficiency laser stripping for a 10 μs, 1 GeV $H^-$ beam. This represents a factor of 1000 increase in the pulse duration compared to the POP experiment, using the same average laser power. The experiment was designed, built, and installed over a three year time period and executed during the spring of 2016. The outcome of the experiment was briefly reported in a Letter [7]. This follow-on paper provides a detailed account of the experimental implementation and all results. The paper is organized into five major sections: (II) implementation of the laser power-saving techniques, (III) predicted stripping efficiencies based on simulations, (IV) experimental configuration and hardware, (V) experimental procedure, and (VI) results.

## II. IMPLEMENTATION OF LASER POWER-SAVING TECHNIQUES IN THE SNS ACCELERATOR

The SNS beam is a 1 GeV, $H^-$ beam produced from a combination of a warm linac (186 MeV) and a superconducting linac (1 GeV). The superconducting linac (SCL) is followed by a high energy beam transport (HEBT) line which includes a 90° bend. The $H^-$ beam has a complicated, multiscale temporal structure, shown in blue in Fig. 2. Macropulses are produced at a rate of 60 Hz; each macropulse is 1 ms in duration and is composed of 1060, 650 ns minipulses; finally, each minipulse contains a train of 50 ps micropulses that repeat at 402.5 MHz. This pulse configuration accommodates the target, ring, and linac cavity structures. Figure 2 also illustrates the pulse structure of the ion and laser beam for this experiment (purple) and for the previous POP experiment (green).

There were three categories of laser power-saving techniques utilized in this experiment: temporal structuring of the laser pulse, manipulation of the ion beam size and divergence, and dispersion tailoring. The techniques and their practical implementation in the SNS accelerator are described here. (i) *Temporal matching of the laser pulse to the $H^-$ pulse structure.*—A substantial laser power savings can be realized by matching the laser pulse structure to the ion beam structure, such that laser power is not spent when there is no ion beam present. Achieving this required moving from the $Q$-switched laser in the POP experiment to a mode-locked laser capable of providing a variable temporal structure. For this experiment, the laser was configured to produce 35–50 ps pulses at the ion beam repetition rate of 402.5 MHz, for a 10 μs duration burst. The temporal matching results in a factor of 70 reduction in the average laser power requirement. (ii) *Dispersion*

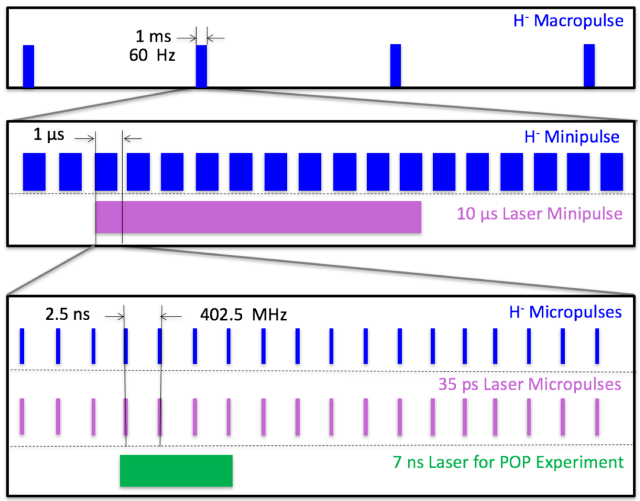

FIG. 2.   The macropulse (top), minipulse (middle), and micropulse (bottom) structure of the SNS H⁻ beam (blue), the laser in this 10 μs experiment (light purple), and the laser in the POP experiment (green). This figure is reproduced from Ref. [7].

*tailoring of H⁻ trajectories.*—Recall that, in the POP experiment, the diverging laser beam was used to address the issue of excitation frequency spread. While this was effective, it was not economical from the laser power standpoint, because a diverging laser beam requires a higher peak power to maintain the required photon density for high efficiency stripping. It was later realized that the excitation frequency spread could be addressed in a more fundamental fashion. Consider the Doppler equation, which relates the frequency of the laser in the lab frame to the frequency of the laser in the rest frame,

$$f_{\text{beam frame}} = \gamma[1 + \beta \cos(\alpha)]f_{\text{lab frame}} \qquad (1)$$

where $\gamma$ and $\beta$ are the relativistic factors of the ion beam and $\alpha$ is the angle of intersection between the laser and the ion beam. Conceptually, the idea is to introduce a correlation between $\gamma$ and $\alpha$ such that each pair produces the same $f_{\text{beam frame}}$. This is done by utilizing the dispersion function which correlates a particle's trajectory with its energy. Mathematically, the requirement of zero frequency spread at the interaction point is given by the relation $D' = -[\beta + \alpha \cos(\alpha)]/\sin(\alpha)$, derived in Ref. [8]. For the SNS 1 GeV beam, with a 355 nm laser placed in the horizontal plane of the ion beam and $\alpha = 39.7°$ for a 1 GeV beam, then $D' = -2.57$ rad. Eliminating the frequency spread in this manner reduces the necessary divergence of the laser and provides an additional factor of 10 savings in the laser peak power. (iii) *Optimization of H⁻ beam size and divergence.*—Following the adjustment of trajectories to reduce the excitation frequency spread due ion beam energy spread, the remaining excitation frequency spread is due to transverse divergence of the ion beam. The angular spread is proportional to $\Delta x' = \sqrt{(1 + \alpha_x^2)/\beta_x}$, where $\alpha_x$ and $\beta_x$ are the horizontal Twiss parameters. Therefore, it is ideal to have $\alpha_x = 0$ and $\beta_x$ as a large value. Finally, the efficiency of the electron excitation to the photon density, and therefore it is beneficial to focus the laser beam to as small a size as possible in order to maximize the photon density for a given laser power. Thus, to produce a high efficiency excitation of the $H^0$ beam, the longitudinal and transverse beam size should be minimized. Ideally, the vertical transverse rms beam size should be <1 mm, and the longitudinal rms microbunch size should be $\sigma_l < 12.5$ ps to guarantee a full overlap of the 50 ps laser pulse. This provides another factor of 2–5 in laser power savings.

The dispersion tailoring and transverse Twiss parameters were achieved by first accurately measuring the Twiss values in the HEBT and then varying nine upstream quadrupoles in a global optimization package that utilizes the XAL online accelerator model [9,10]. The solution is degenerate, and the final choice of optics was determined by testing a few solutions and choosing the one that resulted in the most ideal beam parameters. The same optics set was restored for each experiment, but fine-tuning was required each time to reestablish the exact parameters, especially the vertical beam size at the interaction point (IP). Figures 3 and 4 show the dispersion function measured with the beam position and phase monitors (BPMs) and the typical measured vertical beam size for the experiment using the wire scanner.

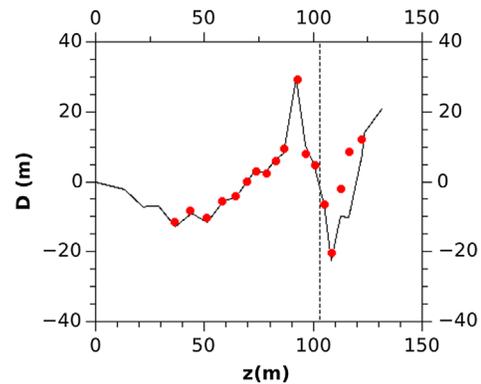

FIG. 3.   Measured dispersion function with dispersion tailoring. The red dots are the measured points, and the black curve is the model prediction. The dashed vertical line indicates the location of the IP. The position $s = 0$ corresponds to the beginning of the first dipole in the HEBT ninety-degree arc.

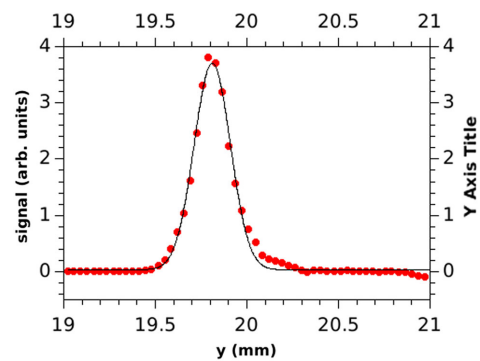

FIG. 4.   Measured vertical profile at the IP.

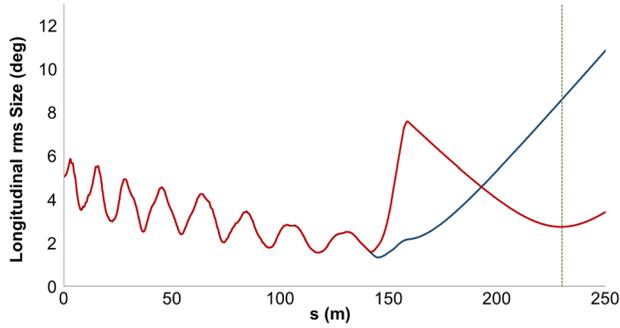

FIG. 5. Longitudinal beam size for production configuration (blue line) and the laser stripping configuration (red line). The dashed line indicates the position of the IP. The position $s = 0$ corresponds to the beginning of the superconducting linear accelerator.

TABLE I. Ion beam design parameters at the IP.

| Parameter | Value |
|---|---|
| $E$ | 1 GeV |
| $D$ | 0 |
| $D'$ | −2.57 |
| $\sigma_l$ | <12.5 ps |
| $\sigma_v$ | <0.1 mm |
| $\alpha_x$ | 0 |
| $\alpha_y$ | 0 |
| Laser peak power | ≥1 MW |

The longitudinal beam size minimization was accomplished by manipulating the phase and amplitude of the last ten SCL cavities. In particular, the first six cavities were used to defocus the beam, the next two cavities were left off, and the next two cavities were used to create a strong focusing kick at the IP. The last cavity was left for fine-tuning the bunch size minimum at the IP using the beam shape monitor (BSM), which measures the longitudinal microbunch profile upstream of the IP. Figure 5 shows the longitudinal beam size for the nominal production beam and for the laser stripping configuration with the minimum bunch length at the IP. Figure 6 shows the BSM measurements for the production beam configuration and the squeezed beam (laser stripping) configuration.

Table I summarizes the design ion beam and laser parameters at the IP of the laser and ion beam that result in high efficiency (≥90%) stripping for a 10 μs, 1 GeV $H^-$ beam.

## III. SIMULATION RESULTS OF PREDICTED STRIPPING EFFICIENCIES

One of the challenges of the laser stripping experiment is that the efficiency of the stripping is sensitive to both laser and ion beam parameters. As a result, the experiment requires the control of ion beam parameters to levels beyond what is necessary for the production operation of the accelerator. Maintaining this control was one of the difficulties of the experiment, as will be discussed later in this document.

To assist with designing the experiment and understanding parameter dependencies, a robust computational model of the laser stripping process has been developed [11]. This model describes the evolution of the $H^0$ beam in the realistic experimental environment where there is a superposition of laser and external magnetic fields. It considers such effects as Stark splitting of the energy levels (and hence the excitation frequencies) and spontaneous emission. The model was applied to the parameters in Table I. Figure 7 shows the overall predicted laser stripping efficiency as a function of the laser peak power for the design parameters. The sensitivity of the stripping efficiency to other important laser parameters is shown in Figs. 8 and 9. Figure 8 shows the dependance of the stripping efficiency on the incident angle of the laser to the ion beam which determines the laser frequency in the rest frame. As seen, the resonance for excitation of the inner electron is very narrow, and a precise angular alignment is required. Figure 9 shows the stripping efficiency dependence on the laser beam size and divergence.

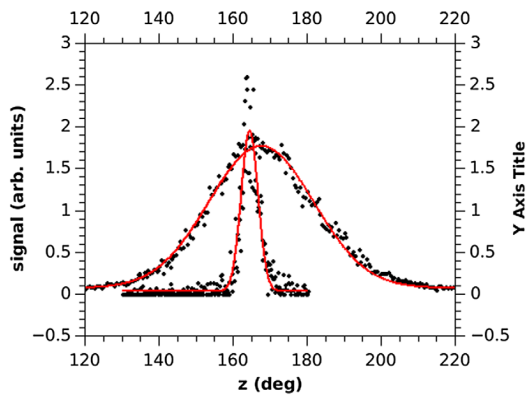

FIG. 6. BSM-measured longitudinal profile for the production beam configuration (wide profile) and the laser stripping configuration (narrow profile). Here, $1° = 3.45$ ps.

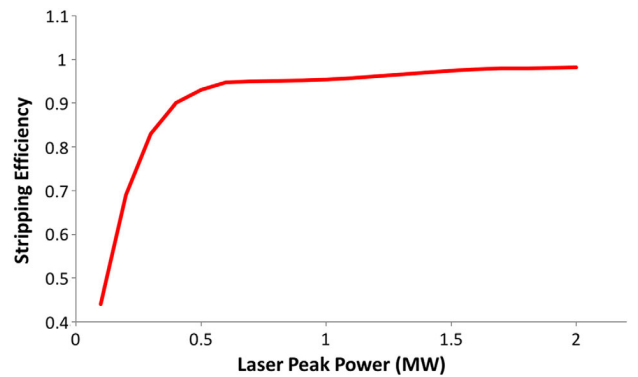

FIG. 7. Predicted stripping efficiency versus beam power for the design parameters.

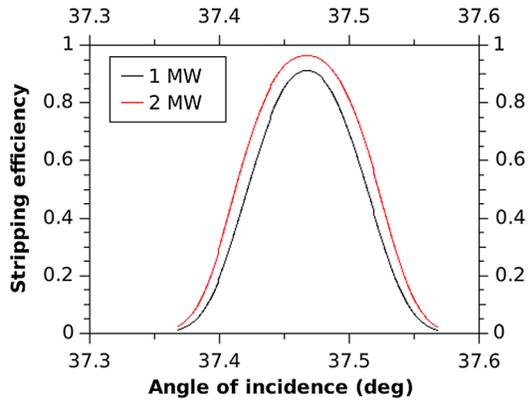

FIG. 8. Predicted stripping efficiency versus laser beam incident angle for laser peak powers of 1 and 2 MW.

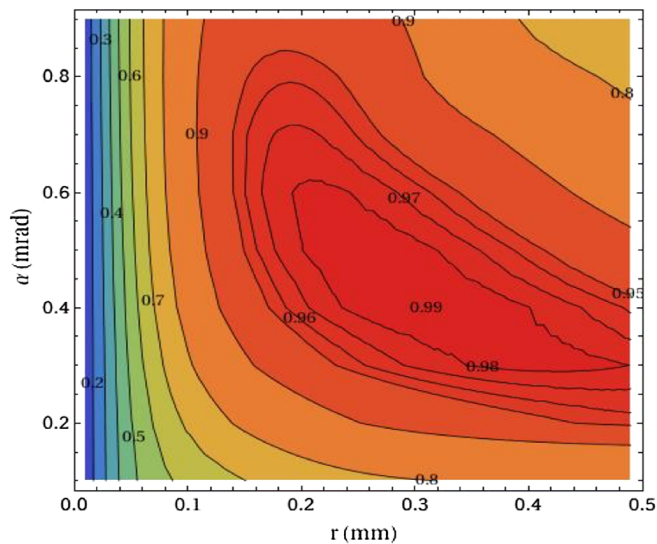

FIG. 9. Map of the stripping efficiency versus laser divergence ($\alpha$) and laser radius ($r$).

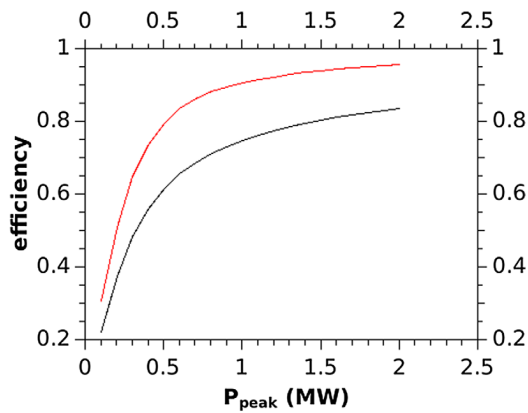

FIG. 10. Laser stripping efficiency with (red curve) and without (blue curve) dispersion tailoring.

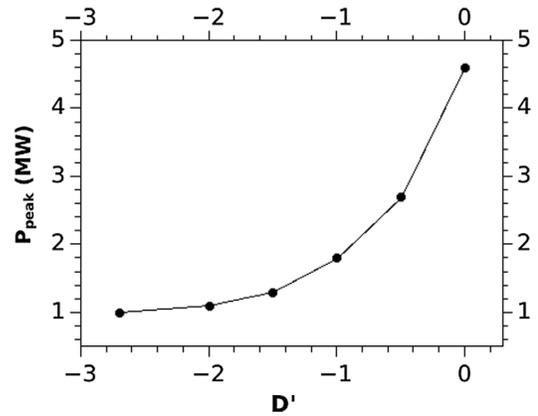

FIG. 11. Peak laser power required for 90% stripping efficiency for different values of $D'$.

It is important to understand the effect of dispersion tailoring, because it is a fundamentally different approach to compensating for the resonant frequency spread compared with the POP experiment. Figure 10 shows the stripping efficiency with and without the dispersion tailoring for laser peak powers up to 2 MW. Although the dispersion tailoring is only an ~15% effect, this corresponds to a dramatic difference in the required laser peak power needed to obtain the higher efficiency curve. To elucidate the point, Fig. 11 shows the laser peak power required to obtain 90% stripping efficiency for various values of $D'$. As shown, obtaining 90% stripping efficiency without the use of dispersion tailoring requires a factor of 5 more in laser peak power, even with other beam parameters as listed in Table I.

## IV. EXPERIMENTAL CONFIGURATION AND HARDWARE

The laser stripping experiment had to be retrofitted into the existing accelerator infrastructure in a manner that optimized success of the experiment while minimizing the project costs and the impact on existing accelerator operations. Five principle objectives guided the design of the experiment: (1) The location must provide the optics flexibility to achieve the ion beam manipulations previously described. (2) The experiment cannot interfere with production operations. (3) The laser system must be protected from radiation damage. (4) The configuration must provide operational flexibility for tuning and multiple experimental attempts. (5) The final design must be capable of achieving >90% stripping efficiency.

A fundamental choice that determines many of the system parameters is the quantum number for the laser resonant excitation of the electron. Here, the choice of $n = 3$ utilized in the POP experiment is persevered, because it represents the best compromise between the lifetime of the excited state, the required stripping field, and the laser power needed for high efficiency excitation. The laser wavelength necessary to excite the electron to the $n = 3$ state is 102.6 nm in the rest frame of the ion beam. The corresponding laser wavelength in the lab frame depends on the beam energy and incident angle as given by Eq. (1). Generally, laser stripping benefits from a higher ion beam energy, because this results in longer wavelength lasers where technology is more developed. The lower limit of ion beam energy for laser stripping is about 1 GeV, corresponding to a UV laser with a wavelength of 355 nm. This experiment was therefore designed for the full 1 GeV H$^-$ beam and, thus, was located downstream of the SCL in the HEBT line between the SCL and the target. The beam exiting the stripping experiment was transported to the downstream ring injection dump.

The dispersion tailoring requires a particular value of the lattice functions $D$ and $D'$, which necessitates that the IP be located downstream of the HEBT arc. It also requires several independently powered upstream quadrupoles to manipulate the dispersion and Twiss functions. For these reasons, the IP was located ~40 m downstream of the exit of the arc, at a location demonstrated to have large optics flexibility. One shortcoming of the chosen IP location is that it is ~150 m downstream of the last longitudinally focusing element. The beam debunches longitudinally over this distance due primarily to space charge effects. Consequently, the minimum achievable rms bunch length increases about ~3.45 ps per mA of beam current. This limitation would not be present in an operational system where a focusing cavity could be placed just upstream of the IP. However, for the configuration at hand, this effect limited the stripping experiment to a few mA of beam current.

To prevent interference with neutron production, the experiment had to be transparent to the beam when not in operation. For this reason, the stripping magnets needed to be either turned off or retracted when not in use. Because of the higher costs associated with electromagnets, retractable permanent magnets were chosen.

Finally, in order to protect the laser from radiation damage and to provide schedule flexibility, the laser was placed remotely in the ring service building and transported in a normal pressure pipe ~70 m to a local optics table adjacent to the IP.

The detailed descriptions of the stripping magnets, the experimental vessel, and the laser system are described next.

## A. Stripper magnets

Focusing the ion beam to a small size at the interaction point, as desired by the power-saving methods previously described, requires preserving the small ion beam emittance in the process of the first electron detachment in the magnetic field. In addition, the magnetic field strength must be low at the IP for an efficient photoexcitation process: less than 20 G for the SNS experimental parameters. The electron has a finite lifetime of the order of nanoseconds in the excited state, and therefore the excited ion must reach the second magnet as quickly as possible, before the electron returns to the ground state. The magnet system for the experiment must satisfy all the above requirements, fit into the available space in the existing SNS beam line, and be compatible with the nominal SNS operation when a high power beam passes through the same beam line.

As will be shown below, the magnetic field of the first magnet must have a gradient of the order of 40 T/m in order to preserve the ion beam emittance. The aperture of the beam line at the experiment location of 0.15 m must be clear for high power operation. A magnet with the field of about 40 T/m · 0.15 m = 6 T at the poles is required to achieve such a large gradient in such a large aperture. A less expensive solution is adopted here with movable magnets of a smaller aperture which are inserted for the stripping experiment and retracted for the normal SNS operation. The whole magnet system has to reside in a vacuum to allow a quick configuration switch without entering the beam tunnel. As the experiment is designed for a fixed ion beam energy of 1 GeV, the magnetic system with a fixed magnetic field strength can be used. This allows the use of permanent magnets instead of electromagnets with a significant reduction of overall system size and cost.

### 1. Magnetic field requirements

There is a finite probability for the electron to detach from an H$^-$ ion in an electric field due to the tunneling effect. The lifetime $\tau$ of an H$^-$ ion in a uniform electric field of strength is

$$\tau[s] = \frac{a}{E[V/m]} e^{\frac{b}{E[V/m]}}, \tag{2}$$

where constants are $a \simeq 2.5 \times 10^{-6}$ sV/m and $b \simeq 4.5 \times 10^9$ V/m [12].

An ion moving in a transverse magnetic field with velocity $v$ will experience an electric field $E = \gamma v B$ in its rest frame of reference and will neutralize with the probability described by Eq. (2). The stripping process is probabilistic, and therefore some ions move in the magnetic field for longer durations than others and thus accumulate larger deflection angles. Thus, the stripping process introduces an angular spread to the ion beam. Seemingly, the angular spread can be reduced by increasing the magnetic field strength, but in a real magnet the field will increase continuously from zero to maximum in the gap and the angular spread will accumulate in the edge field of the magnet. This is illustrated in Fig. 12, where the stripping probability for a 1 GeV H⁻ ion is plotted versus the distance traveled inside a linearly increasing magnetic field. The location of the maximum of the probability function is almost independent of the magnetic field gradient. The probability function width is inversely proportional to the field gradient. The angular spread at the magnet exit versus the field gradient is plotted in Fig. 13 for different lengths of the field ramp. An angular spread of about 1 mrad due to the stripping magnet is a reasonable choice for the interaction point parameters given in the previous sections. Then the required minimum magnetic field gradient is ~40 T/m at the point where the magnetic field strength is 1.2 T.

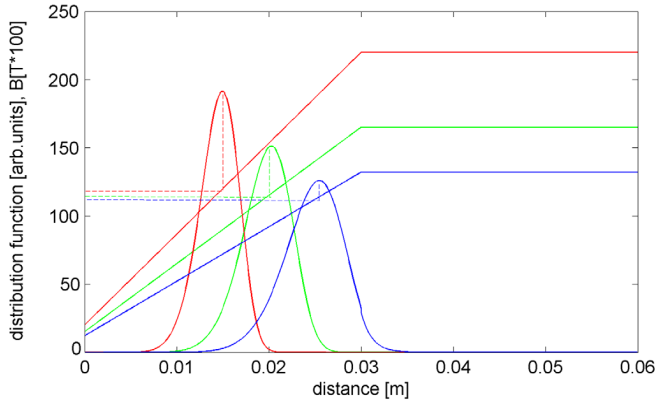

FIG. 12. Distribution of the stripping probability along the beam path (bell-shaped curves) inside a magnet with a linearly ramping fringe field (the magnetic field profile is shown on the same graph).

The requirements for the second magnet that strips the remaining electron excited by the laser are similar to the first one. The stripping process does not depend on the direction of the transverse magnetic field; therefore, it is convenient to have the second magnet identical to the first one with the opposite direction of the field. In this case, the magnetic fields of the two magnets cancel each other at the midpoint between the magnets, and the condition of zero field strength at the IP is satisfied automatically.

The distance from the IP to the second magnet must be as small as possible to minimize the number of excited electrons relaxing to the ground state before they reach the second magnet. On the other hand, it has to be large enough to allow an unobstructed passage of the laser beam through the IP at the required angle of ~40°.

The H⁻ ions are deflected in the edge field of the first magnet before the first electron is detached. Then the protons, created when the second electron is stripped, receive additional deflection in the same direction in the second magnet. In order to minimize the average deflection of the beam exiting the magnets, two dipole field correctors are added before and after the stripping magnets. The field strength of the first corrector must be low enough not to cause premature stripping of the entering ions.

### 2. Magnet design

The Halbach cylindrical array configuration is chosen for the magnets [13], which does not require a heavy and bulky iron yoke for the return field. This results in a compact and light design suitable for actuating inside a vacuum vessel. The magnetic field strength inside the aperture of an infinitely long Halbach cylindrical array with a continuously changing magnetization direction is given by

$$B = B_r \cdot \ln \frac{R}{r}, \tag{3}$$

where $B_r$ is the remnant field of the permanent magnet (PM) material, $R$ is the outer radius, and $r$ is the inner radius of the annular magnet. A typical value of $B_r$ is 1.3–1.4 T for the readily available PM material. The inner radius of 14.5 mm is defined by the ion beam transverse size. The outer radius of 60 mm is chosen to satisfy the magnetic field strength and gradient requirements. Equation (3) gives a very good initial approximation for the magnet parameters. The CST STUDIO SUITE code was used for the final optimization of the design.

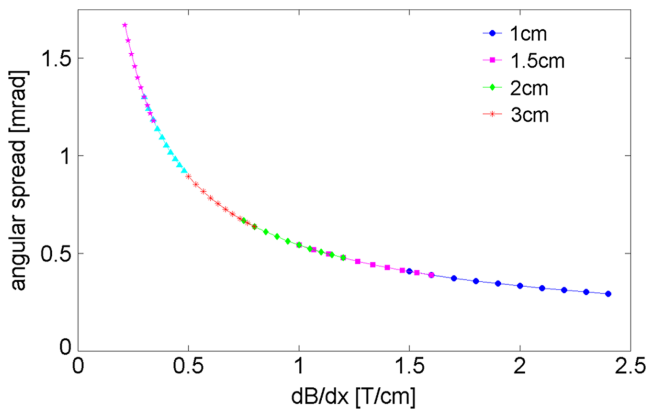

FIG. 13. Dependence of the rms value of the angular spread on the magnetic field gradient (linearly increasing magnetic fields with different maximum field strengths and ramp lengths shown in Fig. 12 were used in the calculations).

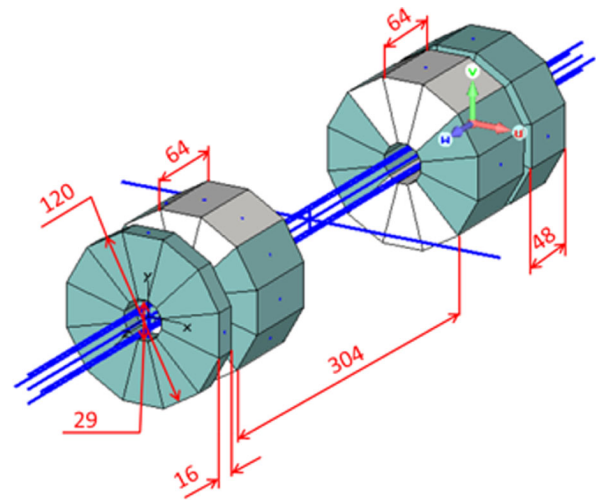

FIG. 14. A layout of the magnet arrangement for the SNS stripping experiment. Dimensions are in millimeters.

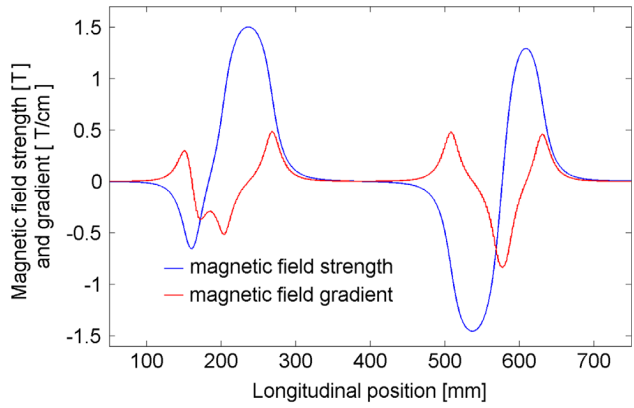

FIG. 15. Plot of the magnetic field strength and gradient distribution along the beam path.

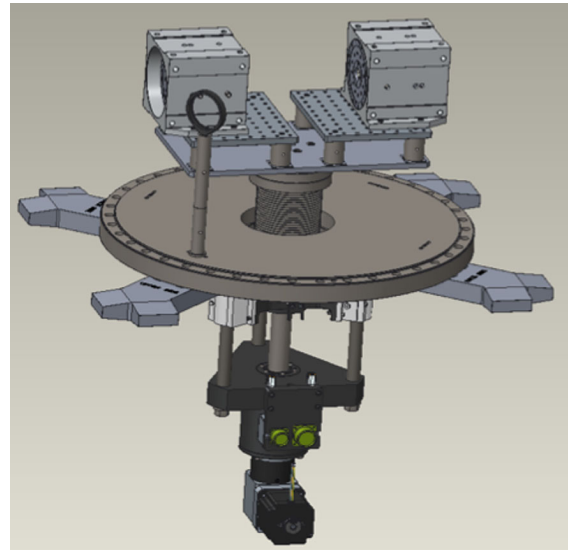

FIG. 17. Magnet system with the actuator assembled on the vacuum flange.

A general layout of the magnet system satisfying all the requirements is shown in Fig. 14. The length of the stripping dipoles is chosen as a compromise between the field strength reduction due to the final length and the total volume of the PM material, which defines the weight and cost of the magnet. The length of the correcting dipoles is chosen to minimize the integral ion beam deflection in the system. The transverse magnetic field strength along the beam path is shown in Fig. 15.

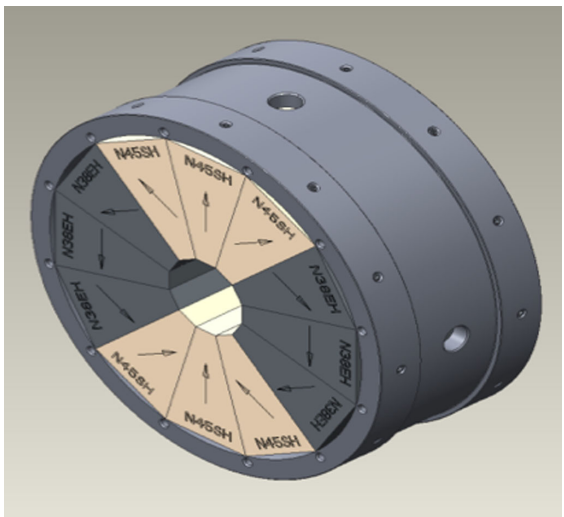

FIG. 16. Model of the individual magnet assembly. Magnetization directions are shown by arrows. Different colors correspond to different PM materials.

Each individual magnet is an assembly of 12 segments held by an aluminum ring as shown in Fig. 16. Two types of PM material are used. The sectors with magnetization collinear with the magnetic field are made of a higher PM material (Dexter N4520 [14]) to maximize the field in the gap. The sectors with magnetization opposite to the magnetic field direction are made of a higher initial coercivity PM material (Dexter N3830 [14]) to prevent PM material demagnetization in high field areas. The magnets were manufactured and measured by SABR Enterprises, LLC [15].

The stripping and corrector magnets in close proximity will attract with a significant force of a few hundred kilograms; therefore, they are mounted in a strong aluminum holding block. The two blocks with pairs of magnets were then mounted on a platform in a vacuum vessel, which is movable up and down using a 200 mm stroke actuator as shown in Fig. 17.

## B. Experimental vessel and diagnostics

A layout of the interaction point experimental set up is shown in Fig. 18. It consists of a vacuum chamber with a magnet system mounted on a remotely controlled actuator, a wire scanner, a beam current transformer (BCM), and vacuum windows for the laser beam entrance and exit. The only dedicated diagnostics used in the experiment are the interaction point wire scanner and the beam current transformer.

The wire scanner is an actuator equipped with horizontal and vertical tungsten wires of 200 $\mu$m diameter. The actuator moves at a 45° angle through the interaction point, thus allowing a measurement of the horizontal and vertical ion beam profiles in one stroke. The wire scanner signal is generated by the electrical charge intercepted by the wires.

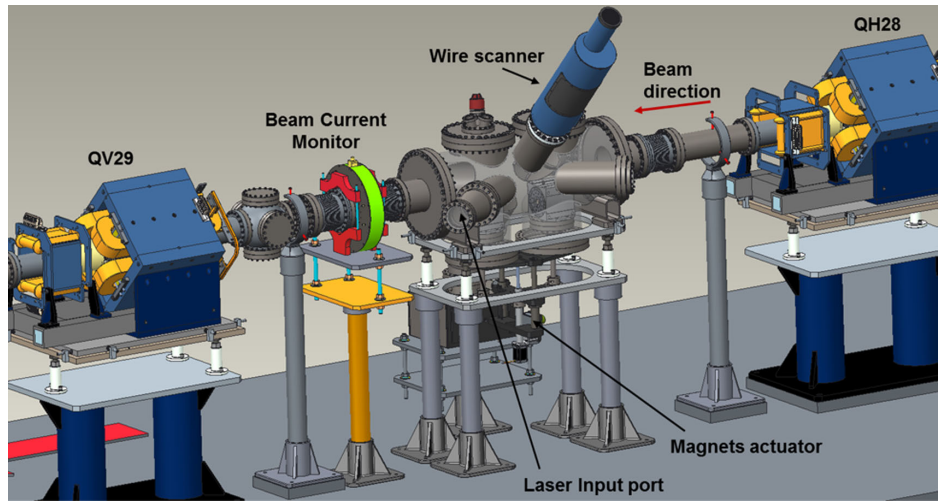

FIG. 18.   A layout of the interaction point experimental setup of the laser stripping experiment.

The wire scanner can also sense the laser beam at the interaction due to thermal electron emission induced by high power density UV light striking the wire.

Both unstripped and stripped ions pass through a current transformer (BCM) immediately downstream of the interaction point. This allows suppressing the systematic calibration error in measuring the stripping efficiency as the unstripped and stripped beams pass through the same detector.

The control and signal processing electronics are installed outside of the tunnel, hundreds of feet away, to avoid damage from the radiation during normal SNS accelerator operation. As a result, significant noise is added to the weak BCM signal. Therefore, a separate high-sensitivity narrow band system measuring the amplitude of 402.5 MHz bunch harmonic is used to measure the ion beam current with much lower noise. As we had only one system installed upstream of the interaction point, only the incoming unstripped ion beam current was measured with low noise.

The baseline SNS diagnostics are used to set up the required beam optics upstream of the interaction point: the wire scanners for the transverse beam size, the BPMs for the dispersion and dispersion derivative, and the BSMs for the longitudinal beam size [16].

## C. Laser system

The overall laser system is composed of the remotely located high power UV laser, the transport line from the remote station to the IP, and the local optical table adjacent to the IP. These systems are described next.

### 1. Remote laser station

The remote laser station uses a master oscillator power amplifier scheme to produce the necessary laser power and temporal structure [17]. It consists of a master oscillator, a pulse picker, a three-stage Nd:YAG amplifier, and harmonic converters as shown in Fig. 19. The master oscillator (seed laser) is an actively mode-locked fiber laser pumped by 1480 nm diode lasers in ytterbium-doped fiber as a gain medium. The laser wavelength is controlled via a temperature-tuned fiber Bragg grating and stabilized at $36.0 \pm 0.05°C$ so that the corresponding wavelength of 1064.45 nm maximizes the Nd:YAG amplifier gain. The pulse width is tunable within a range of 55–85 ps [18].

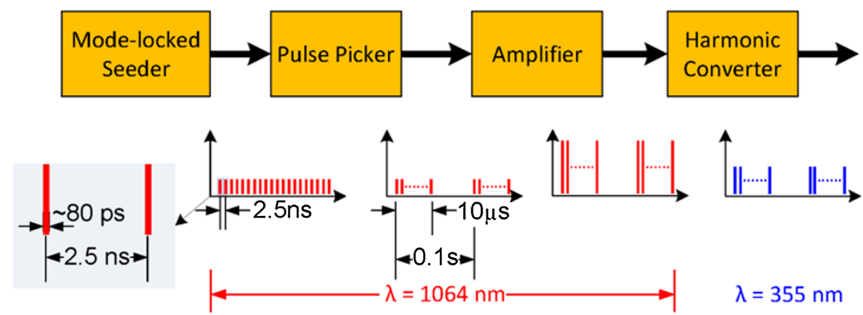

FIG. 19.   Block diagram of the burst-mode laser system illustrates the micro- and macropulse trains and their amplification and conversion processes.

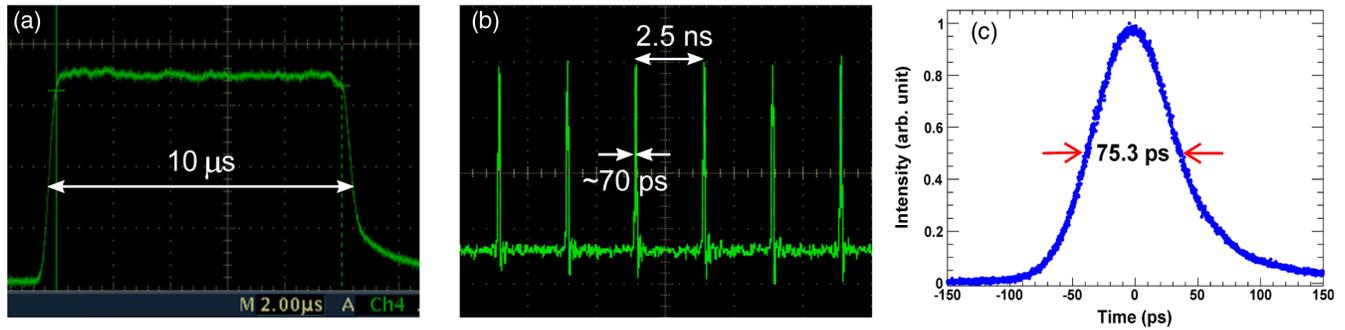

FIG. 20. (a) 10 μs macropulse waveform, (b) micropulse train, and (c) an individual micropulse measured with a fast photodiode.

The seed laser has a built-in fiber amplifier which produces an average output power of 200 mW.

The infrared (IR) pulses from the seed laser are sent to an acousto-optic modulator based pulse picker system that selects 10 μs bursts at 10 Hz repetition rate and subsequently amplifies them in the flash-lamp pumped three-stage Nd:YAG amplifier to produce high energy macropulses. The pulse picker system has the capability of generating macropulses with adjustable pulse durations from submicroseconds to a few tens of microseconds at a repetition of 10 Hz. The macropulse shape is controlled by arbitrary waveforms created on a computer to compensate gain depletion in the Nd:YAG amplifier so that a flat macropulse is obtained at the end of amplification. Figures 20(a) and 20(b) show typical micropulse and macropulse waveforms of the IR beam measured using a photodetector with the peak power of ~2 MW, respectively.

The three-stage Nd:YAG amplifiers pumped by flash lamps provide 6 orders of magnitude power amplification to the input macropulse, and the amplified IR light is subsequently converted to its third harmonic by two lithium triborate crystals [17]. The pulse width of the UV beam has been characterized using a multifunctional optical correlator [18]. For a 10 μs macropulse, the maximum measured peak power of the UV pulses is 3.5 MW at a (micro)pulse width of 33 ± 1 ps. The UV beam quality is limited by the incoming beam size to the Nd:YAG rods in each amplifier and the entire amplification factor (particularly the setting of the last amplifier). The laser beam shows a TEM00

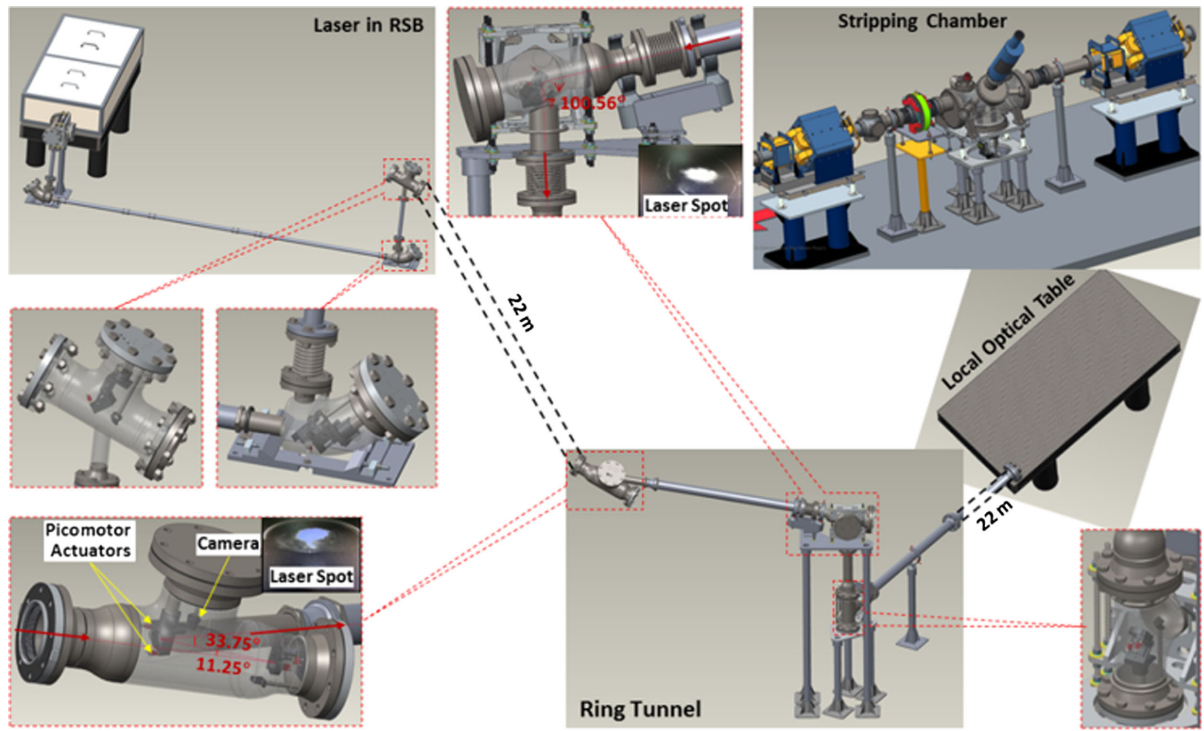

FIG. 21. Schematic of laser transport line. The remote station is located in the SNS ring service building, and the laser beam is transported through a 20 m chase to the ring injection area in the accelerator tunnel using mirrors M1–M8. The optics boxes contain turning mirrors, picomotors, and laser beam position monitoring cameras.

Gaussian mode in the far field with an $M^2$ factor of ~1.3 in both the horizontal and vertical directions.

### 2. Transport line

The SNS accelerator routinely operates at around 1 MW proton beam power. At this power level, the radiation doses in the ring injection area [19] reach tens of kilorads. In order to protect the laser from radiation-induced damage, the laser system was remotely placed in the ring service building (RSB), and the laser beam was transported to the local optics table next to the experimental vessel. The RSB is about 10 m above the beam line and shielded from the accelerator tunnel through a concrete wall. The only available optical path between the RSB and tunnel is a 15 cm penetration chase which was originally designed as a cable chase. A schematic of the laser transport line (LTL) is shown in Fig. 21. It consists of three parts: The first part transports the laser beam from the laser table to the entrance of the penetration hole in the RSB, the second part is a 20-m-long penetration chase, and the last part relays the laser beam from the exit of the penetration chase to the stripping chamber in the tunnel. The entire LTL is enclosed in an atmospheric-pressure aluminum pipe and entrance and exit of the pipe are sealed with 4 inch viewport windows that are antireflection coated at 355 nm. The enclosure is important to eliminate air flow caused by pressure or temperature differences between the RSB and accelerator tunnel. A total of eight mirrors are used to relay the beam in eight different planes along the total beam path of 60 m. All mirrors are 3 inch dielectric mirrors with high-reflection coating at 355 nm over a wide range of incidence angles (5°–53°). All relay mirror mounts except the first two are equipped with a pair of picomotor-driven actuators for remote beam steering and a compact analog camera for beam position monitoring. Many conventional laser transport lines employ the image relay approach where the beam is focused and collimated between two relay mirrors. This approach turns out to be impractical in the present environment due to the limitation of space for lens mounts and lack of control and accessibility. Instead, we chose to propagate a collimated laser beam with diameters of about 10–12 mm ($4\sigma$) through the entire LTL without an image relay. The transmission efficiency was measured to be 70%. Beam losses come from the absorption and scattering on the mirror surfaces, absorption in the air, and limitation of optical apertures during the propagation through the LTL.

### 3. Local laser station

The laser stripping chamber is located ~20 m upstream of the SNS ring injection area and downstream of the superconducting linac [19]. Figure 22 shows a schematic of the local laser station in the SNS accelerator tunnel. The local laser setup consists of a remotely controlled steering mirror, two cameras, a telescope, two steering mirrors, and a power meter at the exit of the experimental vessel. The telescope consists of a lens pair ($f1 = -100$ mm and $f2 = 200$ mm), and the spacing between the two lenses is controlled by a translation stage equipped with a stepper motor. The telescope controls the beam size and divergence angle at the IP. Limit switches are implemented at each end of the stepper motor to restrict the beam sizes on the vacuum windows to prevent laser-induced damage on the vacuum windows. As described in the previous section, successful stripping requires a narrow parameter range in laser beam size, divergence angle, and interaction angle between the laser and ion beams. The interaction angle is controlled by a steering mirror and a second stepper motor. The laser beam is crossing the H$^-$ beam at 37.5°. The entrance and exit windows are located at ~1.0 m from the laser beam focus by using extension tubes. Laser beam positions are monitored before and after the stripping chamber via two Gigabit ethernet cameras. The images acquired by the cameras are sent to the remote computer through the ethernet network for laser beam position diagnostics. A feedback control is applied to the piezo-electric transducer mirror mounted right before the LTL on the remote laser table to reduce the slow (<1 Hz) drifts of the laser beam. Finally, the laser power at the end of the stripping chamber is measured by a power meter to determine the actual laser pulse energy in the stripping chamber. Table II summarizes the major laser parameters measured before the stripping experiment. At the IP, the laser beam has a full divergence angle of 2.4 mrad and a $4\sigma$ beam size of 1.1 mm. Most of the laser operation and laser parameter tuning is remotely controlled in the central control room. The phase difference can be remotely tuned using a computer-controlled digital phase shifter with a precision of 0.1° (corresponding to ~0.7 ps), and the phase is also computer controlled within an accuracy of ~1.0 ns. A full description of the laser system configurations can be found in Ref. [20].

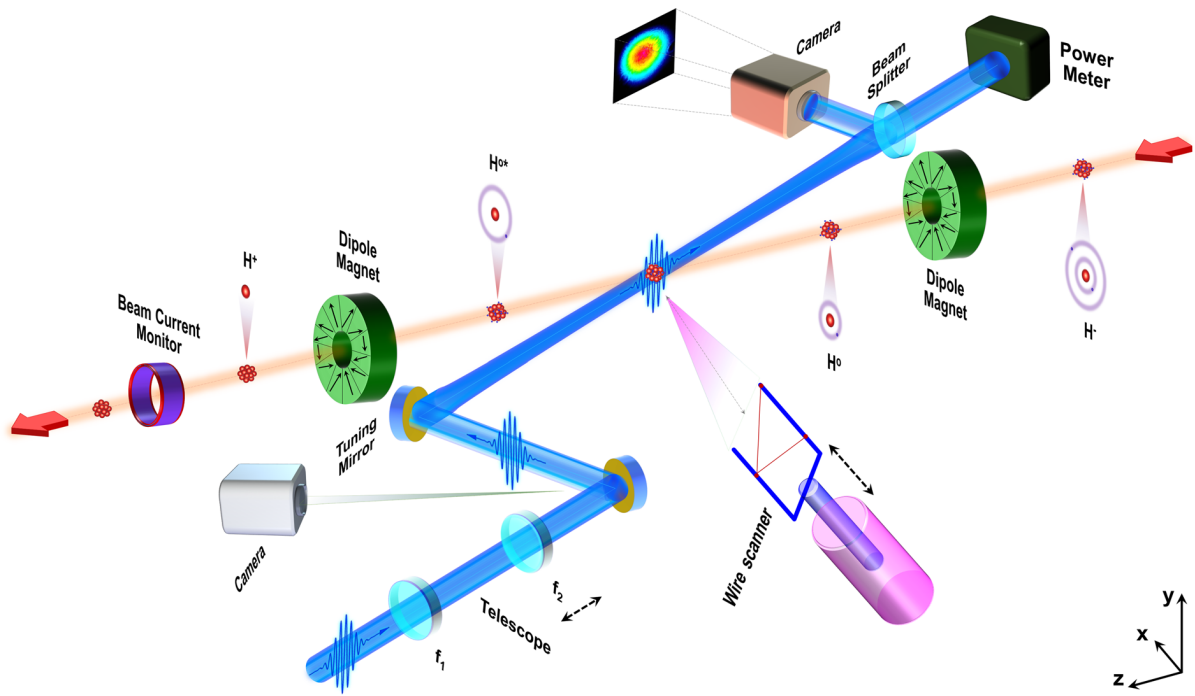

FIG. 22. Schematic of the experimental layout shows three-step charge exchange injection concept along with the laser and ion beam configurations and the instruments for diagnostics during the experiment.

TABLE II.    Table of required versus achieved laser system parameters.

|  | Required | Delivered |
| --- | --- | --- |
| Macropulse length ($\mu$s) | 10 | 12 |
| Micropulse length (ps) | >30 | 30–50 |
| Laser transport line efficiency (%) | >60 | 70 |
| Maximum peak power at IP (MW) | >1.0 | 2.0 |
| Vertical beam full divergence at IP (mrad) | 1.6–2.6 | 2.3–3.0 |
| Vertical beam size ($4\sigma$) at IP (mm) | 1.0–1.4 | 1.1 |
| Pointing stability at IP (mm) |  | $\pm 0.10$ (H), $\pm 0.11$ (V) |
| Max peak power density on vacuum window entry (MW/cm$^2$) | <100 | 57.9 |
| Max peak power density on vacuum window exit (MW/cm$^2$) | <100 | 79.6 |

### D. Final measured parameters

The final measured parameters for the laser system are given in Table II.

## V. EXPERIMENTAL PROCEDURE

All of the hardware for the experiment was installed and tested by early March 2016, and experiments began shortly after. Setting up the experiment was a lengthy process that required multiple teams of system experts during different stages. The experimental setup time to achieve a good laser stripping signal was typically in excess of 12 h. The detailed steps of the experimental procedure were as follows: (1) *Establishing a* ≥ 1 GeV *ion beam energy.*— The operational energy of the SNS linac during the time of the experiment was 957 MeV, limited primarily by field emission in the high beta superconducting cavities [21]. Because the experiment was designed for approximately 1 GeV, it was necessary to tune the beam up for ≥1 GeV, allowing some margin for the manipulation of the last few cavities to achieve the longitudinal beam squeeze. This was accomplished by dropping the SCL rf repetition rate from 60 to 10 Hz and increasing selected cavity gradients by amounts determined through historical commissioning data for each cavity. Fine-tuning of the gradients and cavity heaters was performed to provide dynamic load compensation and long-term operational stability at the lower repetition rate and higher gradient settings. Once stable operation was achieved, the SCL was tuned up to ~1 GeV by appropriately setting the phase of each cavity through a standard SCL tune-up procedure [22]. (2) *Ion beam current and macropulse length.*—The beam current was decreased to a few mA by inserting an aperture in the medium energy beam transport section. The macropulse duration was set for ~10 $\mu$s. (3) *Longitudinal beam squeeze.*—The longitudinal bunch squeeze was performed according to the prescription previously described. This resulted in an approximately 80 MeV energy loss for a final beam energy of 982 MeV. The longitudinal beam size was measured on a downstream BSM and iteration with the last SCL cavity phase, and the amplitude was performed to arrive at the minimum beam size. (4) *Transverse beam optics.*—A set of wire scanners were taken to determine a set of Twiss parameters. Based on this measurement, an online model of the accelerator was used to determine the quadrupole settings that would produce the design optics for 982 MeV. This was done during the first experimental campaign, and thereafter the optics settings were restored and then fine-tuned. (5) *Laser timing adjustment.*—Two levels of temporal synchronization were performed. The micropulses generated in the seed laser were synchronized to the 402.5 MHz rf timing of the accelerator master clock, and the laser macropulse was synchronized to the accelerator macropulse according to a BPM near the stripping chamber. (6) *Vertical position alignment.*—When the laser is incident on the $H^-$ beam, it induces photodetachment of the outer electron yielding $H^0$. The $H^0$ particles are lost in the accelerator generating a detectable beam loss monitor signal downstream of the IP. This process was used to assist with the positional alignment of the laser. First, the laser frequency was offset slightly from 402.5 to 402.51 MHz. This induces a phase sweep such that every 10 s the laser is guaranteed to have roughly the correct microbunch phasing. The wire scanner was used to determine the relative positions of the laser and ion beam center, since they each generate a signal when they intercept the wire. The ion beam was then manually steered with an upstream corrector magnet until the photodetachment beam loss signals on the downstream beam loss monitors (BLMs) were maximized, indicating the optimum positional alignment. (7) *Phase alignment.*—Once the vertical alignment was complete, the laser frequency was restored to 402.5 MHz. The phase of the laser was then scanned from −180° to 180° while monitoring several downstream BLMs to measure the photodetachment signal. The peak of the beam loss was taken as a coarse determination of the correct phase. (8) *Insertion of stripping magnets.*—With all parameters coarsely set, the laser was turned off and the stripping magnets were inserted. Lorentz stripping of the first electron to produce $H^0$ was confirmed by the null signal on the downstream BCM, compared with a few mA on an upstream BCM. (9) *Optimization of laser-ion beam crossing angle.*—The frequency of the laser in the ion beam rest frame is dependent on the ion beam energy and the incoming angle through the Doppler relation [Eq. (1)]. The stripping is sensitive to the crossing angle as shown in Fig. 8, and the only confirmation of the correct setting is the stripping event itself. In this step, the laser was turned on to a relatively low peak power of 0.2–0.3 MW, and the angle was varied in fine steps (∼0.05°) until a stripping signal was confirmed. Because of the positional jitter of the laser and the high sensitivity of this parameter, this was a very lengthy process lasting several hours in some cases. (10) *Fine-tuning.*—Finally, once a verifiable stripping signal was achieved, the vertical steering, laser phasing, laser divergence, and laser angle were finely adjusted until the stripping signal was maximized. (11) *Measuring the stripping efficiency.*—Because of the positional jitter in the laser, it was necessary to gather high data statistics in order to measure the stripping efficiency. Once the stripping was optimized, the stripping magnets were removed, and a baseline set of ∼100 consecutive H⁻ pulse waveforms was measured on BCM28. The magnets were then reinserted, and a set of ≥100 waveforms of the stripped pulses (protons) was measured on the same BCM. If more fine-tuning was done, then the magnets were removed and another H⁻ reference set was taken.

## VI. RESULTS OF THE EXPERIMENT

The laser stripping experiment was performed five times over the course of a three month period. The first goal of the experiments was to achieve high efficiency stripping, and the second goal was to study the dependence of the stripping on key parameters. For reasons that are still unclear, difficulties were encountered in reproducing identical beam optics on each experimental attempt. In addition, the BSM resolution was declining steadily over the months, making the longitudinal bunch length minimization difficult. This was later found to be caused by a dying amplifier in the BSM. This lead to an overall trend of decreasing stripping efficiency over the course of the three months. The experiment was generally restricted to operate with ≤1 MW of laser peak power to prevent any damage to the vessel window that would cause a vacuum leak. However, during the last experiment which occurred just prior to an extended accelerator maintenance period, the full 2 MW of laser power was used.

### A. Data analysis procedure

While some real-time analysis of data was done during the experiment, the final stripping efficiencies were determined by an offline analysis of the BCM28 waveforms. A macropulse consists of a train of individual minipulses, each of which can have a slightly different amplitude due to ion source current transients. Therefore, it is necessary to calculate the stripping efficiencies individually for each minipulse, such that the first minipulse of protons is compared only with the first minipulse of H⁻, and so forth. The laser stripping efficiency is defined as

$$\text{efficiency} = \frac{(I_{\text{proton}})_i}{(I_{H^-})_i}, \qquad (4)$$

where $i$ represents the $i$th minipulses in the macropulse. For the reference H⁻ data set, the analysis was done as follows: For each minipulse in the minipulse train, the data points from the noise floor on either side of the minipulse were averaged to yield an average noise floor level for that minipulse. This value was then subtracted from each point on the flattop of the minipulse to approximately eliminate any noise floor offset. The noise-subtracted data points were then averaged to give a single amplitude beam current value for each minipulse. Finally, the amplitude values were averaged over all shots of data for each $i$th minipulse. This gave a low noise reference H⁻ beam current value for each minipulses in the macropulse train.

For the stripped beam signal, the data analysis was the same up to the point of averaging over multiple shots of data. In this case, since the positional jitter in the laser beam caused a significant shot-to-shot variation in the amount of the stripped beam, an averaging for the proton minipulses was not valid. Therefore, a separate stripping efficiency was calculated for each minipulse in each shot of data. For instance, for a data set with 100 shots of a 10 μs minipulse train, there were $100 \times 10 = 1000$ stripping efficiencies calculated.

### B. Sources of measurement error

There are four significant sources of error in the stripping beam measurement: the H⁻ shot-to-shot current variation, the floor offset and noise on BCM28, and the positional jitter of both the laser and ion beam. These are discussed separately. (i) *H⁻ shot-to-shot current variation.*—A statistically significant set of H⁻ beam current data was gathered using a low noise beam position monitor upstream of the interaction point. From these data, the shot-to-shot variation in H⁻ current was determined to be ≤2%. This error can be practicably eliminated by averaging over several shots of data. (ii) *Positional jitter.*—The positional stability of the ion beam, both within a 10 μs pulse and shot to shot, was measured using an upstream BPM and determined to be insignificant. However, the laser positional jitter is on other orders of the vertical beam size at the IP. Therefore, it is possible for the laser to fully intercept the ion beam on one shot while completely missing it on the next. Additionally, jitter can occur over the duration of a single macropulse. The laser positional jitter is strictly an artifact of the remote placement of the laser and is not related to the physics of the stripping process in any way. A laser system designed for operational use would avoid this problem by colocating the laser with the IP. For the purposes of this experiment, however, the jitter was factored out by gathering statistics over hundreds of minipulses and identifying the largest stripping beam current signals as those with adequate laser and ion beam overlap. (iii) *BCM28 noise and offset.*—A statistically significant data set with no beam was used to determine the noise floor offset and level for BCM28. The analysis indicated no average offset of the noise floor. The noise level was on the order of ~10% of the beam currents use in this experiment, e.g., 0.1 of 1 mA. This is the dominant source of error for the stripping measurement.

### C. Stripping efficiency results

The highest efficiency stripping was observed on the second experimental campaign on March 28, 2016. The BCM measurement of the stripping is shown in Fig. 23. Here, the averaged reference H⁻ beam signal is shown in blue, and eight separate shots of the stripped beam signal are shown in red. In this experiment, a 1 MW peak power, 10 μs laser was used to strip an 11 μs H⁻ macropulse containing 11 minipulses. The laser was aligned with the front side of the macropulse train, such that the last minipulses in the 11 μs macropulse was not intercepted by the laser and therefore remains as H⁰, as indicated by the null signal on the BCM28. The slight decrease in the stripping efficiency near the end of the macropulse is likely due to a small shift in the laser frequency which has been observed in the laser lab.

The data set from which Fig. 23 is contains 456 shots of beam, each with 11 minipulses. Therefore, according to the data analysis procedure previously described, 5016 individual stripping efficiencies were calculated. Figure 24 shows the distribution of calculated stripping efficiencies for the entire data set. The distribution profile is directly correlated with the laser jitter. The peak of the distribution represents cases with good alignment between the laser and ion beam. Since the goal of this experiment is to measure the efficiency of the stripping process when the beams are optimally aligned, the stripping efficiency is determined by the peak value of the distribution, e.g., the highest calculated stripping efficiency. This value is 95%, with an uncertainty of 10%, due to noise in the BCM.

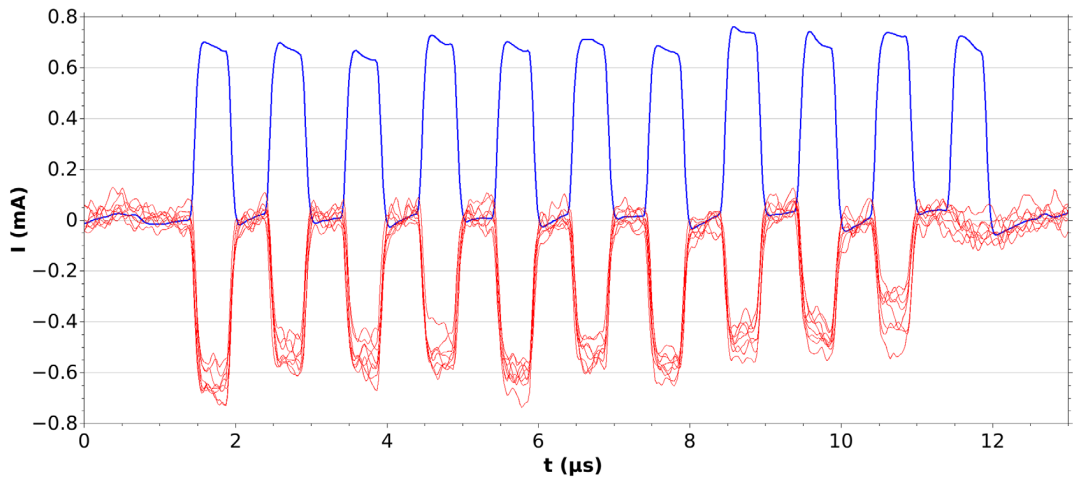

FIG. 23. The experimental results. The average beam current for the H⁻ pulse measured by the BCM at the interaction point before stripping (blue curve) and eight separately measured stripped proton beam shots on the same beam current monitor during stripping (red curve). This figure is reproduced from Ref. [7].

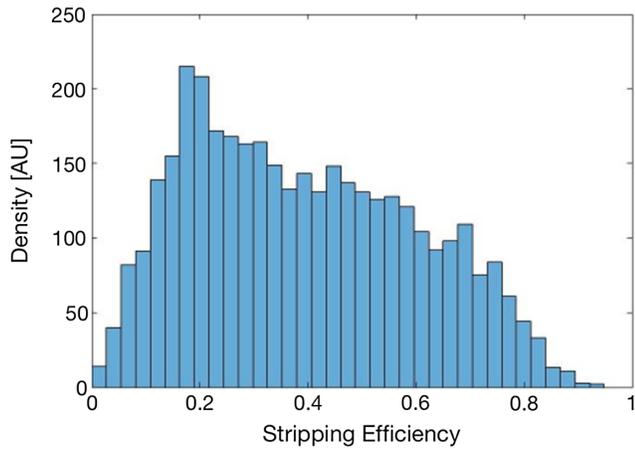

FIG. 24.  Distribution of stripping efficiencies, showing the impact of the laser jitter.

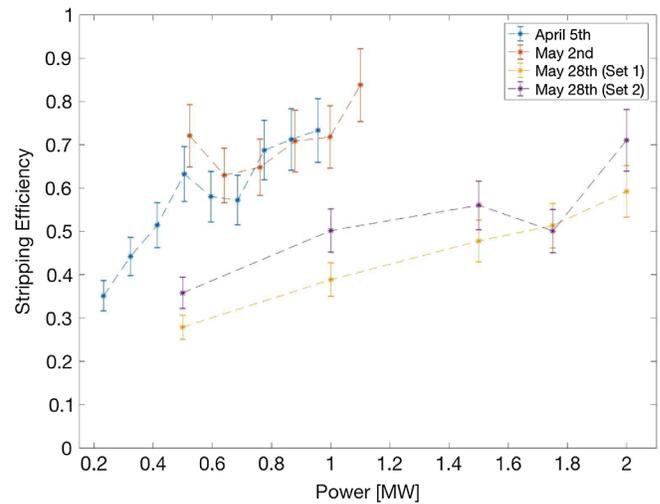

FIG. 25.  Stripping efficiency versus laser power.

Unfortunately, due to the declining resolution of the BSM that limited the longitudinal bunch length minimization and other factors such as time constraints for tuning the transverse optics, the ion beam parameters in this experiment were never precisely reproduced, and subsequent experimental campaigns yielded lower stripping efficiencies in the range of 60%–85%.

### D. Parameter sensitivities

After establishing high efficiency stripping on March 28, the next goal of the experiment was to measure the stripping efficiency versus the key parameters of laser power and $D'$.

#### 1. Laser power dependence

The laser power sensitivity measurement is straightforwardly accomplished by first establishing a strong stripped beam signal, then scanning the laser peak power setting, and gathering a high statistics data set at each setting. The scan was completed four times on three different dates, as shown in Fig. 25. The last measurement (May 28) took place just before an extended maintenance outage of the accelerator, and the allowable peak laser power limit was raised from 1 to 2 MW. The power scan results indicate a strong dependence of stripping efficiency on the laser peak power, as expected. However, none of the scans show a saturation of stripping efficiency with laser power. This was perhaps due to the imperfect settings of the ion beam optics.

#### 2. $D'$ dependence

The $D'$ sensitivity measurement involved a more elaborate process than the laser sensitivity measurement, because each value of $D'$ requires a different set of lattice optics to give the design ion beam parameters at the IP. The process of retuning the lattice optics was time consuming and did not reproduce precisely the same ion beam parameters at the IP for the different $D'$ settings. Since the stripping efficiency is very sensitive to these parameters, this may have affected the results.

The $D'$ sensitivity measurement was performed on May 2. Because of time constraints, only two values of $D'$ were measured: $D' = -2.57$, corresponding to optimum dispersion tailoring, and $D' = 0$, corresponding to no dispersion tailoring and nominal production state. The result is shown in Fig. 26 below. The impact of ideal dispersion tailoring is approximately a 29% effect. This is larger than anticipated from the simulation but may be partially due to more ideal beam parameter tuning in the tailored dispersion state.

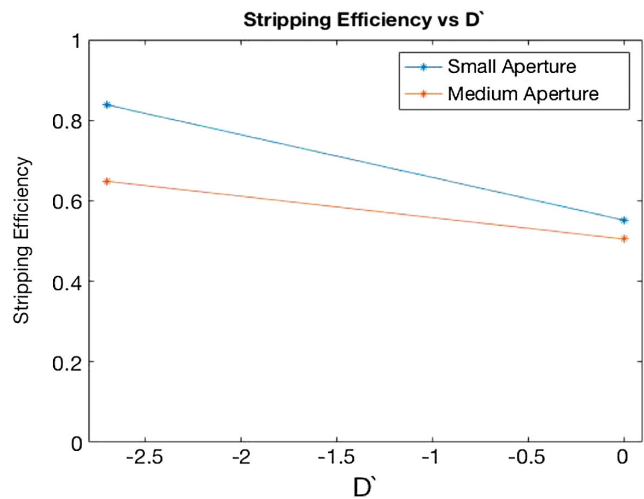

FIG. 26.  Stripping efficiency with and without dispersion tailoring. The laser beam was 1 MW peak power.

## VII. DISCUSSION AND PATH FORWARD

The present experiment successfully demonstrated that laser stripping for a 10 μs H− beam can be achieved with high efficiency (∼95%) using available laser technology. The achieved efficiencies are comparable to the foil-based approach of about 95%–98%. The dispersion tailoring was demonstrated to be effective in compensating for the resonant frequency spread in the ion beam. The main challenges in performing the experiment were in controlling the optics with the necessary level of precision, long setup times for the experiment, and positional jitter in the laser that made the tuning difficult. These problems are all related to the retrofitting of the experiment into the existing SNS infrastructure and would not be present in a laser stripping system designed for production use. The SNS accelerator is expected to reach its nominal 1 GeV beam energy in the winter of 2018, a result of improvements in cavity gradients from *in situ* plasma processing [21]. This will eliminate the need for a SCL reconfiguration prior to the experiment and significantly reduce the experimental setup time.

The final step in the experimental R&D evolution of laser stripping is to demonstrate millisecond-capable stripping, which requires yet another step in laser power savings. The direct scaling of average laser power to full cycle 1 ms, 60 Hz H− beam stripping requires a factor of 600 increase (i.e., ∼1.2 kW average power) from the 10 μs, 10 Hz case while maintaining at least 1 MW peak power. This is beyond the capability of current state-of-the-art burst-mode laser systems. However, since the cross section for the photon-particle interaction in the laser stripping process is extremely small, the stripping event results in a negligible laser power loss (∼$10^{-7}$). Thus, the laser power can be recycled and enhanced if the interaction point is located inside an optical cavity, providing a path forward for full pulse length stripping. Such external optical cavities have been routinely applied to recycle the power from single-frequency lasers or mode-locked lasers which have pico- or femtosecond pulses repeating at tens of megahertz to gigahertz. However, this experiment requires a burst-mode laser with very small duty factors (∼6%). In such cases, it is difficult to generate an effective error signal within the short duration of the laser burst, and the conventional cavity locking technique is not suitable. As part of the development of the next step of this experiment, a different locking method using a doubly resonant optical cavity scheme is being developed to realize a cavity enhancement of burst-mode laser pulses [18]. By using diode-pumped solid-state amplifier and burst-mode optical cavity enhancement technology, it is promising to produce millisecond-long pulse lengths with 1 MW peak power and therefore to demonstrate high efficiency laser stripping of full millisecond duration H− macropulses.

The laser stripping method overcomes the present limitations in foil-based charge exchange injection associated with radiation and foil survivability. For a production-type laser stripping system, the only radiation expected is from the disposal of the 5% partially stripped H⁰ beam via the traditional method of passing it through a thick foil and guiding it to a dump. For 1.2 MW beam operations at SNS, the radiation levels resulting from this process are ≤25 mrem/h at a distance of 30 cm from the thick foil, whereas the radiation from the primary stripper foil that would not be present in the laser stripping system is ∼1 rem/h at 30 cm. The elimination of this high level of radiation would add a new degree of freedom to design strategies for high power accelerators that currently place a heavy emphasis on mitigating foil issues [4]. For instance, injection painting schemes could be optimized entirely from the standpoint of maximizing beam power density, minimizing collective effects, or achieving specialized target or luminosity requirements. Furthermore, the laser stripping technique is particularly attractive for future HEP facilities with H− linear accelerators in the GeV range. Because of the benefit of the Doppler shift of the laser light, at ion beam energies of 3 GeV and above, IR lasers can be used. IR laser technology is well developed and economical compared with the UV laser technology required in this 1 GeV stripping demonstration.

## ACKNOWLEDGMENTS

This work has been partially supported by U.S. DOE Grant No. DE-FG02-13ER41967. ORNL is managed by UT-Battelle, LLC, under Contract No. DE-AC05-00OR22725 for the U.S. Department of Energy. Fermilab is operated by Fermi Research Alliance, LLC under Contract No. DE-AC02-07CH11359 with the United States Department of Energy. Notice: This manuscript has been authored by UT-Battelle, LLC, under Contract No. DE-AC05-00OR22725 with the U.S. Department of Energy.